\documentclass[12pt]{article}
\usepackage{a4wide}
\usepackage{amssymb}
\usepackage{graphicx}
\usepackage{natbib}
\bibpunct{(}{)}{;}{a}{,}{;}
\begin{document}

\newcommand{\case}[2]{{\textstyle \frac{#1}{#2}}}
\newcommand{\lP}{\ell_{\mathrm P}}
\newcommand{\md}{{\mathrm{d}}}

{\renewcommand{\thefootnote}{\fnsymbol{footnote}}
\hfill  IGPG--07/5--5\\
\medskip
\hfill arXiv:0705.4398\\
\medskip
\begin{center}
{\Large \bf The Dark Side of a Patchwork Universe\footnote{Contribution to the special issue on Dark Energy by the journal {\em General Relativity and Gravitation}.}} 

\vskip1cm
Martin Bojowald\footnote{e-mail address: {\tt bojowald@gravity.psu.edu}}
\\
\vspace{0.5em}
Institute for Gravitational Physics and Geometry,
The Pennsylvania State
University,\\
104 Davey Lab, University Park, PA 16802, USA
\vspace{1.5em}
\end{center}
}

\setcounter{footnote}{0}

\begin{abstract}
  While observational cosmology has recently progressed fast, it
  revealed a serious dilemma called dark energy: an unknown source of
  exotic energy with negative pressure driving a current accelerating
  phase of the universe. All attempts so far to find a convincing
  theoretical explanation have failed, so that one of the last hopes
  is the yet to be developed quantum theory of gravity. In this
  article, loop quantum gravity is considered as a candidate, with an
  emphasis on properties which might play a role for the dark energy
  problem. Its basic feature is the discrete structure of space, often
  associated with quantum theories of gravity on general grounds. This
  gives rise to well-defined matter Hamiltonian operators and thus
  sheds light on conceptual questions related to the cosmological
  constant problem.  It also implies typical quantum geometry effects
  which, from a more phenomenological point of view, may result in
  dark energy.  In particular the latter scenario allows several
  non-trivial tests which can be made more precise by detailed
  observations in combination with a quantitative study of numerical
  quantum gravity.  If the speculative possibility of a loop quantum
  gravitational origin of dark energy turns out to be realized, a
  program as outlined here will help to hammer out our ideas for a
  quantum theory of gravity, and at the same time allow predictions
  for the distant future of our universe.
\end{abstract}

\section{Introduction}

A complete understanding of the universe currently faces several
problems, most of which are occasionally expected to be solved by some
version of quantum gravity. This also applies to the dark energy
problem. Its first, older part is the cosmological constant problem,
i.e.\ the divergence of vacuum energy density when it is naively
computed in quantum field theory, and its being off by orders of
magnitude if computed with reasonable-looking cut-offs. More
precisely, the question is how vacuum energy couples to gravity, and
which subtraction of the quantum field theoretical result has to be
employed to derive the gravitationally relevant value. Any fundamental
theory must provide a solution to this part of the problem, since a
fundamental theory must provide a well-defined answer to a
well-defined question such as that of the energy density in a vacuum
state.\footnote{Not all general relativistic situations allow a global
energy concept, but at least an approximate notion of energy density
must exist for nearly vacuum states of fields in a maximally symmetric
space-time.} Whether or not quantum gravity can solve this part of the
dark energy problem is thus part of the question whether quantum
gravity will be a fundamental theory. This is sometimes expected, but
there is an independent relation to quantum gravity: an ultraviolet
cut-off would make the vacuum energy finite, which is often done by
hand although it gives, at best, confusing results. Since one often
associates an underlying space-time discreteness with quantum gravity,
and thus an upper bound for wave numbers and frequencies, a natural
cut-off may be provided to result in a finite vacuum energy. A precise
implementation would provide means to determine the correct coupling
of vacuum energy to gravity.

While the second part of the problem---nailing down a dynamical
culprit for dark energy---is more specific, its relation to quantum
gravity is only vague. This part of the problem became acute only
after recent observational indications that the current evolution of
the universe is best modeled by a Friedmann--Robertson--Walker
space-time if a negative pressure component to its energy balance
became dominant in the recent past.
Giving a precise value for the corresponding dark energy contribution,
combined with information on its equation of state, the problem
becomes more specific, and also closer to providing explicit tests of
falsifiable candidates for solutions to the dark energy problem or the
quantum gravity theories on which they may be based. This is, for
instance, possible in candidate theories leading to additional, so far
unobserved, fields which could provide the dark energy component.

Loop quantum gravity \citep{Rov,ALRev,ThomasRev}, the topic of this
article, is not of this kind of a theory with surplus fields, at least
not in its presently realized form. There are no fields in addition to
the gravitational one and whatever matter fields one assumed before
setting out to quantize the theory. (It could, however, be possible
that additional degrees of freedom emerge in a yet to be developed
effective description; see e.g.\ \citep{trinion} for a suggestion.)
What is realized in loop quantum gravity and in fact presents one of
its main features is an underlying discrete structure of space
\citep{AreaVol,Area,Vol2}, so that one may at least address the first
part of the problem. This requires making precise the relation between
spatial discreteness and quantum field theoretical cut-offs. As for
the second part of the problem, what is provided by existing
phenomenological descriptions of some regimes of loop quantum gravity
are modifications to matter equations of state which, in an average
effect, could resemble a dark energy contribution. These
possibilities, which are currently all to be considered speculative,
are described here.

The specific difficulty of addressing the dark energy problem from the
viewpoint of loop quantum gravity is that it requires detailed
information on the behavior of inhomogeneous configurations. While the
dynamics of loop quantum gravity has been understood quite well in
homogeneous models \citep{LivRev}, which already provides access to
several problems in cosmology, this cannot result in deviations from
classical gravity resembling dark energy in large, very classical
universes. Quantum corrections to homogeneous models only appear when
the total spatial size is small, close to the Planck scale, or when
anisotropies are so big that curvature components become large. This
is certainly not realized in those recent phases of our universe which
the dark energy problem refers to. Any dark energy contribution from
loop quantum gravity must then be more indirect; it cannot be a single
quantum correction term to a classical evolution equation but at best
a cumulative effect, obtained by coherently adding up small
corrections.  A precedent for such an effect in loop cosmology has
recently been found \citep{InhomEvolve}: small quantum corrections can
add up coherently during long cosmic evolution times to give more
sizable and potentially observable effects. If something similar
happens in space instead of time, a sizable dark energy component
could result by adding coherently small quantum corrections occurring
at separate spatial points in a nearly Friedmann--Robertson--Walker
spacetime. As we will see below, this provides a well-defined setup
which can be addressed, though not yet fully evaluated, in loop
quantum gravity.

We will first introduce the basic features of loop quantum gravity and
cosmology. This will then be used to review its present contributions
to both parts of the dark energy problem. The first part is being
addressed by an attempt to make contact between quantum gravity, in
which all fields including the gravitational one are quantized, and
quantum field theory on a classical curved manifold. The second part
can be analyzed in the context of specific models which describe how a
discrete quantum state of space-time evolves, including refinements of
the discrete structure as the universe expands. This refers to
detailed dynamical properties, which makes the problem technically
complicated but also indicate that a dedicated analysis can provide
stringent tests of the underlying formulation of loop quantum gravity.
As we will see, some of the known properties of dark energy are
successfully modeled in this way.

\section{The universe according to loop quantum gravity}
\label{s:Ham}

Universe dynamics is described in loop quantum gravity by an
``evolving'' graph as a superposition of states representing the
spatial quantum geometry. Many of the typical properties of physical
scenarios based on loop quantum gravity are directly related to this
underlying discreteness, which will also be the main scheme used in
this article. The emphasis on spatial geometry arises because of the
canonical quantization used to set up the theory. As in classical
canonical formulations, a covariant space-time will result only if
certain constraint equations are satisfied. Solutions to the quantum
constraints, taking the form of superpositions of spatial geometry
states, encode the ``evolution'' of physical fields classically
depicted as a space-time manifold.

\subsection{Spatial geometry}

In a canonical formulation of general relativity \citep{ADM} only the
spatial metric components $q_{ab}$, referring to a chosen slicing of
space-time by a global time function, are dynamical fields and have
conjugate momenta. The remaining components of the space-time metric
tensor play the role of Lagrange multipliers of the constraints
mentioned above, analogously to the time component of the vector
potential in electrodynamics which becomes the Lagrange multiplier of
the Gauss constraint. Accordingly, only spatial geometric objects such
as areas of spatial surfaces or volumes of spatial regions are
represented as operators in a canonical quantization. Loop quantum
gravity proceeds by first reformulating the theory in terms of
densitized triads $E^a_i=\sqrt{\det (q_{bc})}e^a_i$ with the triad
$e^a_i$ satisfying $e^a_ie^b_i=q^{ab}$ canonically conjugate to a
connection $A_a^i$ \citep{AshVar,AshVarReell}. This allows one to
define holonomies
\begin{equation}
 h_e(A) = {\cal P}\exp\int_e\md t \dot{e}^aA_a^i\tau_i
\end{equation}
of the connection $A^i_a$ along curves $e$ in space with ${\cal P}$
denoting path ordering of the non-Abelian SU(2) matrices (with
generators $\tau_i$) along the path, and fluxes
\begin{equation}
 F_S(E) = \int_S\md^2y n_aE^a_i \tau^i
\end{equation}
of the densitized triad through surfaces $S$ with co-normal $n_a$.
These objects, unlike the local fields $(A_a^i,E^b_j)$ themselves,
form a well-defined Poisson algebra without the delta functions one
otherwise has in field theories. This algebra has a representation as
operators on a Hilbert space, which is the basis of loop quantum
gravity.

The representation is most conveniently written in the connection
representation. Then, being functionals of the connection as
configuration variables, states are constructed from holonomies along
paths in space and thus themselves refer to 1-dimensional objects. One
can formulate this by using arbitrary graphs in space as a label of
states, replacing the coordinate position of the classical fields
\citep{RS:Spinnet,ALMMT}. Edges and vertices of the graphs carry
further quantum labels, which for edges are half-integers
$j\in\frac{1}{2}{\mathbb N}$. (The resemblance to spin quantum numbers
is due to the occurrence of the spatial rotation group, acting on the
triad, in this context.) These quantum numbers assigned to a spatial
graph determine the spatial geometry by combinatorial rules. The area
of a surface, for instance, is obtained by summing contributions
$4\pi\gamma \ell_{\rm P}^2 j(j+1)$ for all edges in a graph
intersecting the given surface \citep{AreaVol,Area}, with a
coefficient determined in terms of the Planck length $\ell_{\rm P}$
and the so-called Barbero--Immirzi parameter $\gamma$ of loop quantum
gravity \citep{Immirzi,AshVarReell}. More precisely, this formula
gives the eigenvalue of the area operator when it acts on a state
associated with the given graph and edge labels. In particular, the
area spectrum is discrete. This formula only applies to the generic
case where the surface does not intersect any of the vertices of the
graph. If this were the case, vertex labels would also be necessary to
determine the action of the area operator, but the spectrum remains
discrete. For volume, only vertex labels play a role in a way more
complicated to specify \citep{AreaVol,Vol2}. But the action is
obtained in a similar combinatorial manner by summing contributions
from vertices within a given region whose volume is to be
determined. Moreover, the volume spectrum can also be shown to be
discrete although it is not known explicitly.

\subsection{Relational evolution}

Such a state, associated with a single graph or as a superposition of
states associated with different graphs, describes spatial quantum
geometry at an instant, so far without an evolution or space-time
picture. An arbitrary state does correspond to a physical quantum
space-time only if it satisfies constraint equations, i.e.\ if it is
annihilated by operators quantizing the classical constraints. This in
general requires a physical state to be a superposition of different
volume eigenstates. Evolution is encoded in such a superposition in a
relational manner, for instance as the change of an area or of a
matter field relative to a change in volume: For a positive real
number $V$ we can project a given superposition to the volume
eigenspace with eigenvalue $V$ (which may be the zero space) and
compute the expectation value of an area $A$ or a matter field $\phi$
in the projection. Doing this for all values of $V$ results in a
function $A(V)$ or $\phi(V)$ of volume which describes the relational
evolution. This presents an implicit description of space-time not as
a manifold but as the relational dynamics of the fields defined on it.
We do not refer to $A(t)$ and $V(t)$ as functions of a coordinate time
variable $t$ which, in a general relativistic situation, could be
chosen arbitrarily. We rather eliminate $t$ in regions where $V(t)$ is
invertible by inserting $t(V)$ in $A(t)$. The variable $A$ then
evolves with respect to the ``internal time'' $V$. Just like
coordinates, this description in general can be done only locally in
patches corresponding to the image domain where $V(t)$ is invertible.
To cover the whole dynamics relationally, one will have to select
different internal times and derive transformation rules between them.
(In a triad formulation one can even compute evolution for all real
values of such an internal volume time including negative ones since
the theory is sensitive also to the orientation and not just the size
of space. This is relevant for the singularity issue, which is
resolved in loop quantum cosmology by providing a well-defined
relational evolution through zero volume or other degenerate
configurations \citep{Sing,SphSymmSing,BSCG}.) Evolution in canonical
quantum gravity thus does not happen in coordinate time since there is
no time manifold, but in a relational manner. Observables have been
defined directly in such a way
\citep{BergmannTime,GeomObs1,GeomObs2,PartialCompleteObs,DittrichThesis}.

Just as not any 4-dimensional Lorentzian manifold is a physical
space-time but only those are which satisfy Einstein's equation,
allowed superpositions of states in quantum gravity need to satisfy
the constraint equations. These equations select physical
superpositions of spatial geometry states giving rise to relational
evolution. They are formulated by specifying the moves that can occur
in the underlying graphs and labels of states when an internal time
variable such as $V$ changes. Schematically, one has a picture where
space is presented as a discrete structure building up from a small
state at the big bang to a highly refined, nearly continuous fabric
today.  The evolution picture is thus that of an irregular lattice
structure which changes in internal time by elementary changes of
geometry. Corresponding moves follow from the quantization procedure
applied to the classical constraints. Although there is currently no
uniquely specified version, several typical features of these moves
are generic and can thus reveal properties of the fundamental quantum
dynamics.

\begin{figure}
\begin{center}
\includegraphics[width=14cm]{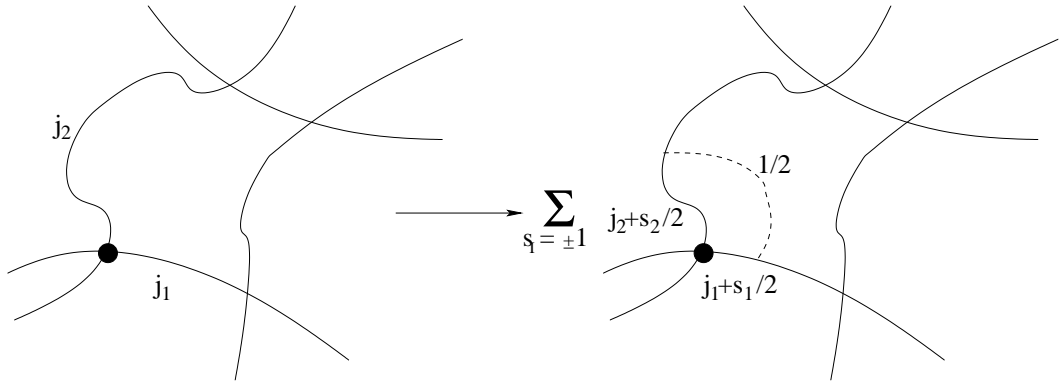}
\end{center}
\caption{Local action of typical Hamiltonian constraint operators on a
  vertex of a graph, indicated by a disc. Edge labels which change
  during the action are specified. \label{local}}
\end{figure}

Unfortunately, a derivation of such rules from first principles, i.e.\
by quantizing the Hamiltonian constraint and reformulating the
constraint equation in terms of an internal time, is highly complex.
Fig.~\ref{local} illustrates the local action of a typical constraint
operator as it follows from the usual constructions
\citep{RS:Ham,QSDI}. The operator, when acting on a single graph
state, changes labels as well as the graph itself, resulting in
several new contributions to the total superposition. (However, also
versions of Hamiltonian constraint operators based on fixed lattices
have been constructed; see e.g.\ \cite{AQGI}.)  Any single change of
this form may increase or decrease the geometrical variable chosen as
internal time, and thus represents elementary changes of the graph
forward or backward in internal time. This is to be reformulated as
global moves of the physical solution forward in time by rearranging
the total solution as a superposition of internal time eigenstates.
Since local moves acting on distant vertices can change the total
internal time (e.g.\ volume) by equal amounts, the resulting forward
moves, unlike the elementary ones in the quantized constraint, may no
longer be local, adding to the technical complexity. We will now
describe the form of the elementary constraint operator and later
formulate basic assumptions on its implications for moves in internal
time.

\subsection{Properties of Hamiltonians}

Properties of elementary moves are consequences of the main terms
appearing in the classical Hamiltonian constraint of canonical general
relativity in Ashtekar variables, quantized by loop techniques. The
main contribution to the constraint is
\begin{equation} \label{Ham}
 H[N] = \frac{1}{16\pi G} \int_{\Sigma} \mathrm{d}^3x N
 \left(\epsilon_{ijk}F_{ab}^i\frac{E^a_jE^b_k}{\sqrt{|\det
E|}} \right)
\end{equation}
which classically must vanish for any spatial function $N$. The
constraint depends on the densitized triad as well as the connection
through its Yang--Mills curvature $F_{ab}^i$. Both dependencies
require special care \citep{RS:Ham,QSDI}: Curvature terms can be
quantized by expressing them as holonomies along closed loops which,
by the non-Abelian Stokes theorem results in the exponentiated
curvature. For a sufficiently small loop, the exponential becomes
irrelevant and one obtains an expression of $F_{ab}^i$ in terms of
quantities acting in the loop representation. Such holonomies act as
multiplication operators and either generate new edges in the graph of
a state acted on (as in Fig.~\ref{local}) or, if the loop is only
retracing edges already present in the graph, change the edge and
vertex labels. In general, one thus expects from this action a change
in the graph and a change in local geometrical values such as area and
volume. Each state in the superposition on the right hand side of
Fig.~\ref{local} in general has a volume expectation value different
from the initial state.  The action thus gives rise to the elementary
moves in internal time mentioned above, which can in principle be
derived once an explicit quantum operator for the constraint has been
specified and physical solutions as superpositions of internal time
eigenstates are analyzed.

Formulating a complete operator also requires one to represent the
triad dependent expression in the constraint. It appears as an inverse
determinant, which classically can diverge and may be problematic in
the quantum theory, too. As we have seen, the volume in loop quantum
gravity acquires a discrete spectrum, and it does contain zero as an
eigenvalue. Thus, there is no inverse operator and no obvious way to
find a quantization of $|\det(E^a_i)|^{-1/2}$ as required for the
Hamiltonian constraint. The second characteristic aspect of a loop
quantization, in addition to loop holonomies quantizing $F_{ab}^i$, is
a consequence of this behavior and is thus directly related to the
discrete spectrum of the volume operator. Rather than taking a direct
inverse, expressions such as that in the Hamiltonian constraint can be
quantized by first rewriting them in a way involving Poisson brackets
between holonomies and volume only requiring positive powers. The
basic relation used is \citep{QSDI}
\begin{equation} \label{ident}
 \left\{A_a^i,\int{\sqrt{|\det E|}}\mathrm{d}^3x\right\}= 2\pi\gamma G
 \epsilon^{ijk}\epsilon_{abc} \frac{E^b_jE^c_k}{{\sqrt{|\det E|}}}
\end{equation}
which follows from the fact that the connection $A_a^i$ is canonically
conjugate to the densitized triad.  In such expressions, all
ingredients can be turned into operators in a well-defined manner, and
eigenvalues of the resulting operator do approach the classical
behavior for large volume and small anisotropy. But at small volume or
for large anisotropies there are deviations from the classical
behavior giving rise to quantum corrections in any classical
expression where an inverse of the densitized triad occurs.

The qualitative behavior of such inverse powers for small anisotropies
can be illustrated well by explicit formulas from isotropic models
\citep{IsoCosmo}. Here, we have a single component $p$ of the
densitized triad $E^a_i=p\delta^a_i$, and a single connection
component $c$ in $A_a^i=c\delta_a^i$ satisfying $\{c,p\}=8\pi\gamma
G/3$. Analogously to (\ref{ident}) we write
\begin{equation}
 \{c,|p|^{1/2}\} = \frac{4}{3}\pi\gamma G |p|^{-1/2}
\end{equation}
showing the resulting inverse power of $p$ even though no inverse is
used on the left hand side. 

\begin{figure}
\centerline{\includegraphics[width=10cm,keepaspectratio]{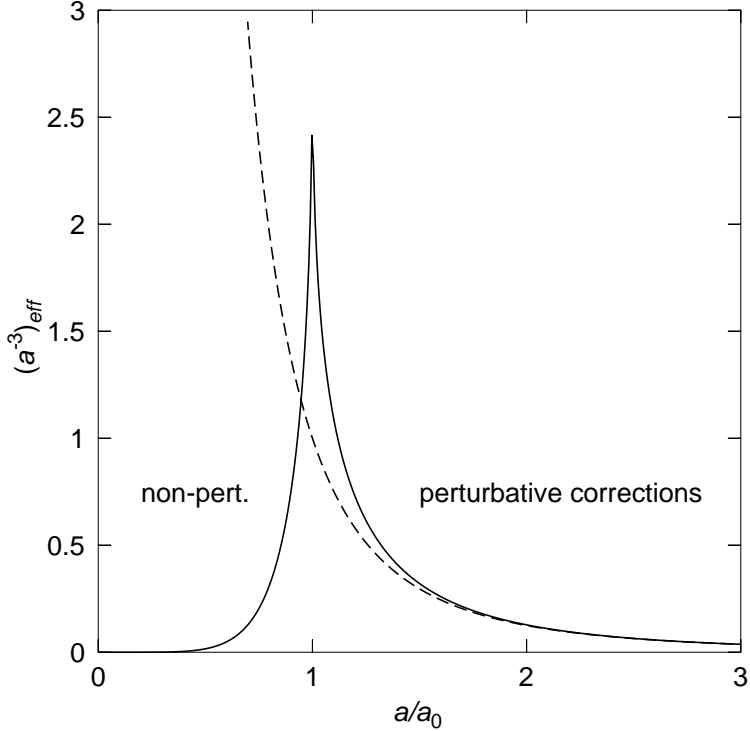}}
\caption{Eigenvalues (\ref{inverseeigen}) cubed of the inverse volume
operator compared to the classical expectation $a^{-3}$ (dashed). To
the right of the peak deviations are due to perturbative corrections
in $\ell_{\rm P}/a$, while at and to the left of the peak stronger
non-perturbative corrections occur. The functions are plotted with
respect to a normalized scale factor $a/a_0$,
$a_0^2=\frac{8}{3}\pi\gamma\ell_{\rm P}^2$, such that the peak occurs
at $a/a_0=1$. \label{inverse}}
\end{figure}

Similarly to holonomies in the full theory, isotropic models of loop
quantum cosmology only allow the quantization of exponentials $e^{i\mu
c}$ on a Hilbert space $\ell^2({\mathbb R})$ by
$(\psi_{\nu})_{\nu\in{\mathbb R}}\mapsto
(\psi_{\nu-\mu})_{\nu\in{\mathbb R}}$, not of $c$ itself
\citep{Bohr}. The Poisson bracket thus can be quantized only after $c$
is expressed in terms of such exponentials, which in the isotropic
case can be done exactly by
\begin{eqnarray}
 \{c,\sqrt{|p|}\} &=&i\delta^{-1}e^{i\delta c}\{e^{-i\delta
 c},\sqrt{|p|}\}\nonumber\\ &=& \frac{i}{2\delta}\left( e^{i\delta
 c}\{e^{-i\delta c},\sqrt{|p|}\}- e^{-i\delta c}\{e^{i\delta
 c},\sqrt{|p|}\}\right)\,. \label{invbrack}
\end{eqnarray}
The parameter $\delta$ is to be chosen as a quantization ambiguity or
may follow from a symmetry reduction of the quantum operator; see
\citep{SymmRed,SymmQFT,Reduction,SymmStatesInt,Rieffel} for these
developments. In the full theory where SU(2) expression are used,
values for $\delta$ are restricted to be half-integers.  In the second
line of (\ref{invbrack}) we have ensured that the expression remains
even under $c\mapsto -c$ (including the derivative by $c$ implicit in
the Poisson bracket), which is required if the phase factors of an
isotropic model are to be embedded as matrix elements of SU(2)
holonomies in the full theory. (The map $c\mapsto -c$ then appears as
a consequence of the Weyl group.)  Note, however, that the full
situation with its non-Abelian holonomies taking values in SU(2) rather
than just phase factors is more subtle \citep{DegFull}, giving rise to
additional quantum corrections compared to models based on Abelian
U(1) holonomies. In isotropic models we then proceed by turning
elementary phase space functions into basic operators and the Poisson
bracket into a commutator divided by $i\hbar$,
\begin{equation} \label{inverseiso}
 \widehat{\frac{1}{\sqrt{|p|}}}= \frac{3}{8\pi\gamma\delta\ell_{\rm
 P}^2} \left(e^{i\delta c}[e^{-i\delta c},\sqrt{|\hat{p}|}]-
 e^{-i\delta c}[e^{i\delta c},\sqrt{|\hat{p}|}]\right)
\end{equation}
(with the Planck length $\ell_{\rm P}= \sqrt{G\hbar}$) which is
densely defined and even finite \citep{InvScale}. We only need to take
the square root of the norm of $\hat{p}$, which for a self-adjoint
operator $(\hat{p}\psi)_{\nu}=
p_{\nu}\psi_{\nu}=\frac{8}{3}\pi\gamma\ell_{\rm P}^2 \nu\psi_{\nu}$
can easily be done. Using the action of the basic operators one can
directly compute the action and eigenvalues of (\ref{inverseiso}) in
explicit form:
\begin{equation} \label{inverseeigen}
 \left(\frac{1}{\sqrt{|p|}}\right)_{\nu}=
\sqrt{\frac{3}{8\pi\gamma}}\frac{1}{\delta \ell_{\rm P}}
\left(\sqrt{|\nu+\delta|}-\sqrt{|\nu-\delta|}\right)\,.
\end{equation}
For $|\nu|\gg\delta$ the expression is close to the classical value,
\begin{equation} \label{inverseeigenexpand}
 \left(\frac{1}{\sqrt{|p|}}\right)_{\nu}= \frac{1}{\sqrt{|p_{\nu}|}}
\left(1+\frac{4\pi^2\delta^2\gamma^2\ell_{\rm
P}^4}{9p_{\nu}^2}+O(\ell_{\rm P}^8/p_{\nu}^4)\right)
\end{equation}
with only perturbative corrections in $\ell_{\rm P}^2/p_{\nu}$, while
there are strong, non-perturbative deviations for $|\nu|$ comparable
to and smaller than $\delta$; see Fig.~\ref{inverse}. The most drastic
deviations occur at $\nu=0$ where (\ref{inverseeigen}) vanishes while
the classical expectation diverges. More general expressions including
quantization ambiguities have been derived in \cite{Ambig,ICGC}. This
illustrates why operators for inverse densitized triad components can
be well-defined in loop quantum gravity and remove divergences in
Hamiltonians which would arise when fields are quantized on a
classical metric background.

In the isotropic example used here, the basic densitized triad
component is related to the conventional scale factor by $|p|=a^2$. A
distinguished value of $a$ defined in terms of the Planck length, such
as the peak position of the quantized inverse, seems to be in conflict
with the classical rescaling freedom of $a$ by an arbitrary positive
factor present in a flat isotropic model. This only poses an apparent
problem which is resolved once operators are viewed in the proper
inhomogeneous context \citep{InhomLattice}. Corrections to the
classical behavior such as those embodied by the peak appear as
functions of the local discrete scales, or local fluxes associated
with individual edges in a graph. These local flux values can be
viewed as elementary building blocks of a macroscopic region. Their
size determines how many building blocks are present in a given region
of a certain volume, and is thus relevant for physical implications:
whether or not discreteness effects are significant depends on the
ratios of elementary flux values to a macroscopic scale. In this way,
the underlying discrete structure distinguishes scales even if the
geometry is scale invariant macroscopically.

Inverse components of the densitized triad also appear in matter
Hamiltonians, such as in
\begin{equation} \label{Hphi}
 H_{\phi}[N]= \int_{\Sigma}\md^3x
\left(\frac{1}{2}\frac{p_{\phi}^2}{\sqrt{|\det E^a_i|}}+ \sqrt{|\det
E^a_i|}V(\phi)\right)
\end{equation}
for a scalar field $\phi$ with momentum $p_{\phi}$. Again, the loop
quantization proceeds by rewriting the classical inverse determinant
in the kinetic term in a form accessible to quantization \citep{QSDV},
giving the classical behavior at large volume but quantum corrections
when the total volume becomes smaller or anisotropies are large.

\subsection{Summary}

Writing the basic action of a Hamiltonian constraint operator through
moves of a discrete graph while an internal time variable such as
volume changes may be difficult. But there are rather robust general
aspects of the picture which we are going to use in what follows.
Hamiltonian operators are discrete, with countably many contributions.
Each contribution receives characteristic corrections to the classical
form, which includes discretization corrections but also quantization
effects such as those seen in (\ref{inverseeigenexpand}) for inverse
powers of the spatial metric determinant. The size and number of all
these contributions depends on the precise state which describes the
universe. All ingredients are dynamical and in general change when an
internal time variable changes. For instance, as the universe grows
one expects the number of contributions to increase, corresponding to
a refined lattice structure.\footnote{This is a general expectation,
  not just based on basic properties of loop quantum gravity. It has
  been discussed in the context of string theory, too, for instance by
  \cite{UniverseExpandString}.} A quantum correction from a single
contribution may grow or shrink with this refinement, the precise
behavior following from the operators.

Even in the absence of a direct line of derivations from a fundamental
Hamiltonian to a refinement model, the form of operators indicates the
general behavior of correction terms and how they change with a
refined graph. The difficult part is to derive how an initial graph
state evolves in internal time and how it is being refined. But once
such a refinement prescription is known or assumed, the behavior of
quantum corrections is rather robust. This makes it possible to
construct phenomenological models based on an assumption for graph
refinements occurring with a change of internal time, which then
implies the internal time dependence of corrections. A
phenomenological Hamiltonian will result which allows one to analyze
potential physical implications of these effects. The robustness of
the procedure is underlined by the fact that there are independent
arguments for certain refinements even in symmetry reduced models
where one can analyze the dynamical equations more easily. Good
semiclassical behavior can then be used as a criterion to select some
refinements and rule out others \citep{APSII,SchwarzN}.

What we will use therefore is a phenomenological Hamiltonian depending
on a function ${\cal N}(V)$ describing the average number of vertices
of a state as a function of the spatial volume. Changes in ${\cal
  N}(V)$ indicate refinements of the graph during evolution. We cannot
compute this function at the current level of developments. It can be
a complicated function, but for short evolution times, which is
sufficient for the dark energy problem, one can assume a power law
${\cal N}(V)\propto V^{\alpha}$. Theoretical considerations indicate
that $\alpha$ should be in the range $0<\alpha<1$: a fundamental
Hamiltonian constraint operator typically creates new vertices and
increases the edge spins. If edge spins remained constant, the growth
in volume could only come from the number of vertices at constant
vertex contributions and thus ${\cal N}\propto V$. If only edge spins
change but the number of vertices remains constant, we obviously have
${\cal N}={\rm const}$. Thus, since both vertex creation and edge spin
increase happens at the same time, the actual power low must be
somewhere in between. Restrictions could be found by detailed
semiclassical analyses as in \citep{SchwarzN}, or phenomenologically
as we will see below.

\section{The dark energy problem: Possible contributions from loop
quantum gravity}

Although loop quantum gravity by its general formalism provides a
well-defined formulation of possible quantum dynamics, including the
gravitational as well as matter fields, the resulting equations for
physical states annihilated by the constraint operators as sketched
above are complicated. They do simplify in symmetric models, most
notably in homogeneous ones as they are often used in cosmology, but
this is unlikely to provide insights to the dark energy problem of a
large, classical universe. All facets of the dark energy problem, from
the viewpoint of loop quantum gravity, require properties of
inhomogeneous states. This involves a more detailed analysis of the
full constraint equations, which is still in progress. Nevertheless,
despite this lack of understanding of details there are indications
for possible contributions of loop quantum gravity to the dark energy
problem.\footnote{We only discuss here mechanisms by which dark energy
  could be provided through an ingredient of loop quantum gravity.
  Investigations in loop quantum cosmology where a fluid with negative
  equation of state parameter is assumed are not covered; see e.g.\
  \citep{FutureSingLoop,PhantomLoop}.}

\subsection{Implications of connection variables}

We start with results which do not, at least currently, use loop
quantum gravity but which crucially rely on the same basic variables
given by real Ashtekar variables. Since many general properties of
loop quantum gravity---the use of holonomies and fluxes as basic
operators of a well-defined background independent quantization with
the resulting discrete flux spectrum---are consequences of the type
of basic variables used, independent implications of these variables
are likely to be realized in loop quantum gravity, too.

The Ashtekar connection $A_a^i=\Gamma_a^i+\gamma K_a^i$ is defined in
terms of the spin connection $\Gamma_a^i$ compatible with the
densitized triad $E^a_i$ and extrinsic curvature $K_a^i$ only up to a
real parameter $\gamma>0$. This constant, the Barbero--Immirzi
parameter \citep{AshVarReell,Immirzi}, can be changed arbitrarily by a
canonical transformation for pure gravity or with bosonic fields. As
pointed out recently by \cite{FermionImmirzi}\footnote{Related
observations were made earlier by \cite{NonPert} and \cite{NPZRev}.} and
elaborated by \cite{FermionTorsion} and \cite{FermionAshtekar}, however, the
value of $\gamma$ does have a physical effect on fermion fields due to
their coupling to the connection. The fermion current provides a
source for torsion.  Solving for and eliminating torsion in the Dirac
action, one obtains an effective current-current interaction whose
strength depends on $\gamma$.

\cite{ImmirziLambda} observe that the form of this four-fermion
interaction resembles what is used in the BCS theory of
superconductivity. Non-perturbative effects imply the instability of
the fermionic ground state and its decay into a condensate effectively
described by a scalar. This condensate may provide a source of dark
energy which is dynamical in the history of the universe. The
corresponding scalar would then play the role of either an inflaton or
quintessence field. Loop quantum gravity may help in formulating the
specifics of this construction but has not been applied yet.

\subsection{Finiteness of vacuum energy}

The construction of Hamiltonian constraint operators and matter
Hamiltonians sketched above results in operators which are
well-defined mathematically. A normal ordering as used in quantum
field theories on curved background space-times, where Hamiltonians
would be diverging without proper subtractions, is not needed. This
makes it possible, at least in principle, to determine the coupling of
vacuum energy to the gravitational field within the theory. Additional
assumptions on the form of subtraction need not be imposed, although
there are quantization ambiguities which may influence the resulting
value of gravitationally active vacuum energy. Despite of the presence
of such ambiguities, a derivation would indicate whether or not a
large amount of fine tuning would be necessary to be in agreement with
current observations.

The finiteness of Hamiltonian operators in loop quantum gravity can be
traced back to the underlying background independence \citep{QSDI}: by
taking into account the full metric in the quantization, rather than
treating it as a classical background field, regularization parameters
which when sent to zero would give rise to divergences in a
quantization on a background cancel out completely in the background
independent quantization. Thus, Hamiltonians are well-defined
operators with dense domains of definition, and any physical situation
can be described by a quantum state with finite expectation values of
Hamiltonians. (Since these operators are typically unbounded, not
every state lies in their domain of definition. But any physical
situation can be described arbitrarily closely by a state in the
domain of definition due to its denseness, which is sufficient given
the finite precision of any measurement.)  In particular, the
expectation value in the vacuum of a quantum field in a quantum
universe must be finite, and the vacuum energy cannot diverge. The
central question is whether or not this just reproduces the value
obtained by choosing a Planck-size cut-off in a Fock quantization.

While these methods lead to such a very general statement about the
finiteness of the vacuum energy, which could play the role of dark
energy if its size has the correct magnitude, specific results are
more difficult to derive. First, one would like to make contact
between this fully background independent quantization and quantum
field theories on a curved background so as to see how the common
divergences are avoided precisely without an explicit normal
ordering. Preliminary results have been derived by
\cite{QFTonCSTI,QFTonCSTII}. From the point of view of quantum field
theory on curved space-times one can effectively view the finiteness
of vacuum energy in loop quantum gravity as a cut-off provided by the
underlying discrete structure of loop quantum gravity. On the grounds
of dimensional arguments one would expect that the cut-off occurs at
Planckian values of energy or length, which would certainly result in
the well-known mismatch between the predicted and observed
cosmological constants. To understand this issue one has to go beyond
dimensional arguments and do a dedicated calculation of expectation
values of Hamiltonians in semiclassical states which can be associated
with a quantum field theoretical vacuum. Here, the full complexity of
loop Hamiltonians strikes and makes the calculation very
involved. Preliminary investigations have been done, but not yet
resulted in any numerical estimates. 

It is to be expected that vacuum energy in this formalism does not
only depend on the matter state but also on quantum geometry, i.e.\
the precise discrete structure emerging from a physical state as
described in Sec.~\ref{s:Ham}. Dimensional arguments can then easily
fail due to the presence of several length scales: the Planck length,
a macroscopic scale such as the Hubble length in cosmology, and the
discreteness scale somewhere in between the former two scales. Loop
quantum gravity in general gives only a lower bound for the
discreteness scale close to the Planck length, but the precise value
can well be larger depending on the actual quantum state. It has to be
determined from properties of a physical solution of quantum gravity
which can describe our universe. For a discreteness scale different
from the Planck length, dimensional arguments are not sufficient for
simple estimates of results and detailed calculations are
required. This is similar to effective theories in a cosmological
context as pointed out in \cite{SILAFAE}. Effective arguments also
apply to the problem at hand here, as described next.

\subsection{Effective negative pressure}

Rather than understanding dark energy as the vacuum energy of quantum
fields, it could be a quantum effect which, when expressed in a
Friedmann--Robertson--Walker solution, resembles an effective matter
contribution giving rise to negative pressure.  In loop quantum
gravity, an explanation of dark energy could be provided in this
manner. In isotropic models, loop quantum cosmology has revealed a
source for negative pressure from quantum corrections
\citep{Inflation}. This happens on small scales where quantum geometry
modifies the behavior of functions such as the inverse volume which
classically diverge at zero spatial extension. As seen in
Fig.~\ref{inverse}, the isotropic quantum version of the inverse cuts
off the divergence and bends the curve down to zero
\citep{InvScale}.\footnote{This is no longer the case for the
  eigenvalues of analogous operators in anisotropic situations.
  However, in that case it is not the full spectrum which is relevant
  but the part of the spectrum realized along a dynamical state (i.e.\
  evaluated in volume eigenstates in its superposition if volume is
  used as internal time) . In isotropic models, both concepts coincide
  since the minisuperspace is 1-dimensional and thus any geometrical
  trajectory reaching zero volume must fill out the whole small volume
  region. This is different in the absence of isotropy where dynamical
  information is much more crucial \citep{DegFull}.}  Effective matter
Hamiltonians, which always contain the inverse determinant of the
densitized triad, are thus increasing as functions of volume at early
times which replaces the classical decreasing behavior. It is then
easy to see why negative pressure arises: by thermodynamics pressure
is defined as the negative derivative of energy by volume. If energy
starts to increase with volume when small scales are reached, negative
pressure automatically results.

\subsubsection{The scenario}

While these effects can be used in inflationary scenarios which
require negative pressure at early times
\citep{Inflation,InflationWMAP,Robust,GenericInfl}, the effects
quickly subside once the universe has expanded to some size very small
compared to its current one. In fact, such a quantum geometry epoch of
inflation typically does not last long enough to provide all 60
$e$-foldings required for successful structure formation. Moreover,
such an isotropic model with only inverse volume corrections is not
very accurate at large volume because it does not fully take into
account the dynamical discreteness of space manifesting itself in
lattice refinements determined by the elementary moves of a
Hamiltonian constraint. Rather, during expansion the discrete
structure of space subdivides as described in Sec.~\ref{s:Ham} which
can be modeled by adding new small, discrete patches resulting from
new vertices of graphs.  The number of patches is expected to grow
with the volume, ${\cal N}\sim V^{\alpha}$ for some constant
$0<\alpha<1$, which indeed has recently been recognized to resolve
some technical problems related to isotropic models on large scales
\citep{APS,InhomLattice,SchwarzN}. The parameter $\alpha$ is to be
determined from the precise behavior as depicted in Fig.~\ref{local}.
When the number of patches increases with volume, their size stays
nearly constant or could even decrease. Thus, some of the small-scale
corrections active in the very early phase can remain or re-emerge
later if they had subsided. How precisely this happens is a
complicated dynamical process of a system with many independent but
dynamically coupled ingredients. While a derivation from first
principles is currently out of reach, one can arrive at estimates on
the overall behavior based on the history of the universe. After the
first, quantum geometry powered phase of inflation the strong effects
causing this phase must subside due to growing edge labels. It follows
a long stretch of evolution with only minor corrections, during which
most of the standard model of cosmology takes place (including a
second phase of inflation). While there are no strong quantum gravity
effects in this regime, small perturbative corrections are always
present and may change some of the observable properties due to their
effect on perturbative inhomogeneities \citep{InhomEvolve}. Such
effects of perturbations around Friedmann--Robertson--Walker
space-times do not contribute negative pressure to the energy balance.
But when inhomogeneous properties of non-perturbative states are
considered, a negative pressure contribution results as we will
show below.

This scenario is in principle testable by direct calculations in the
full theory which would have to demonstrate the sketched behavior of
edge spins: they first have to grow on average when the total volume
is still rather small leading to a semiclassical regime, but then
remain nearly constant (or even start to decrease) while the universe
as a whole expands. Thus, there should be a point where an overall
growth in both edge spins and the number of vertices turns over to
growth dominated by the number of vertices at nearly constant edge
spins.  Independently of whether or not this can be shown from a
fundamental Hamiltonian, one can already test the scenario
phenomenologically.  The question is if this effect will be sufficient
for the amount of dark energy needed, or maybe even too much?

\subsubsection{A phenomenological model}

To determine this more quantitatively one needs to use the relevant
equations as they are corrected by quantum gravity effects. Rather
than dealing directly with Hamiltonian operators and semiclassical
states we write a classical Hamiltonian which includes some of the
effects described previously. This can be seen to arise as an
expectation value of the fundamental matter Hamiltonian in a
semiclassical state, along the lines of a general scheme to derive
effective equations \citep{EffAc,Karpacz}. In particular, we
incorporate the discreteness of space, building the total matter
energy by summing over discrete patches. The size of each patch is
taken to be dynamical to take into account the occurrence of lattice
refinements. Finally, whenever there is an inverse of densitized triad
components in the matter Hamiltonian we include correction functions
for effects seen in quantum inverse operators.  Then, matter energy
with a classical form (\ref{Hphi}) for a scalar field $\phi$ as before
is given by a Hamiltonian\footnote{We just use this as a typical
  example but do not assume the presence of any fundamental scalar
  field; expressions will be similar for realistic matter fields. In
  fact, a strength of the loop quantum cosmological mechanism to
  provide negative pressure is that it would work for any matter field
  through its kinetic term in the Hamiltonian, rather than a
  fine-tuned potential of a postulated scalar. See
  \citep{MaxwellEOS,FermionEOS} for analogous quantum corrections to
  Maxwell and Dirac Hamiltonians.}
\begin{equation} \label{HphiDisc}
 H_{\rm
matter}=\sum_{v}\left(\frac{1}{2}\frac{p_{\phi}(v)^2}{R(v)^3}+
V(\phi)R(v)^3\right)= \sum_{v}R(v)^3\rho_{\rm matter}(v)\,.
\end{equation}
In this expression, we use the patch position
$v$, its radial size $R(v)$ and local energy density $\rho(v)$. The
inverse patch volume $R(v)^{-3}$ present in the kinetic term is the
main term receiving quantum corrections in its effective expression
\begin{equation}\label{da}
 d(R)=R^{-3}(1+CR^{-2}\ell_{\rm P}^2+\cdots)
\end{equation}
with a constant $C>0$ which is determined by the specific
quantization. Here, we are using only perturbative corrections to the
inverse, excluding non-perturbative ones which are realized at much
smaller values of the patch size where the universe would not be
semiclassical at all.  The value of $C$ in inverse volume operators is
subject to quantization ambiguities \citep{Ambig,ICGC} also affecting
the Hamiltonian operator itself from which an effective Hamiltonian
(\ref{HphiDisc}) is derived. The eigenvalue expression
(\ref{inverseeigenexpand}), for instance, used for an expression
$\left(|p|^{-1/2}\right)^3_{\nu(a)}$ resulting in $a^{-3}$ for large
$a$, gives a coefficient $C=4\pi^2\gamma^2\delta^2/3\approx
0.742\delta^2$ which, using the value $\gamma\approx 0.2375$ as
derived from black hole entropy calculations
\citep{ABCK:LoopEntro,IHEntro,Gamma,Gamma2}, is of order one if $\delta$
is of order one as expected; see
\citep{InhomLattice}.\footnote{Although (\ref{inverseeigenexpand}) was
  derived for an isotropic model with $a^{-3}$ being the total inverse
  volume, a very similar expression results for the local patch sizes
  \citep{QuantCorrPert}.} However, in this expansion the leading
correction term is of the order $\ell_{\rm P}^4/p^2=\ell_{\rm
  P}^4/a^4$, and there is no term of order $\ell_{\rm P}^2/p$ since
the quantization is invariant under orientation reversal $p\mapsto
-p$. Terms of second order can nevertheless result when one uses
expectation values in semiclassical states, as they are relevant for
effective equations \citep{EffAc,EffectiveEOM,Karpacz},
rather than eigenvalues. The computation of the corresponding $C$ then
requires more detailed knowledge of the state but can in principle be
performed.

General properties of the quantization imply that $C$ must be
positive, a property which will play an important role. Moreover,
generically only even powers of $\ell_{\rm P}/R$ occur in the
correction terms because the expression for inverse powers is a
function of basic, area-like fluxes, used in the underlying quantum
theory, whose norm is the square of $R$. With this correction, there
will be a new term
\[
 H_{\rm quantum}=\frac{1}{2}\sum_v\left(d(R(v))-R(v)^{-3}\right)
 p_{\phi}(v)^2= \frac{C}{2}\sum_{v}p_{\phi}(v)^2
\frac{\ell_{\rm P}^2}{R(v)^5}
\]
in any matter Hamiltonian $H_{\rm matter}= H_{\rm classical}+ H_{\rm
  quantum}$. This can be estimated using an approximately homogeneous
matter distribution and the virial theorem which states that energy
contributed by the kinetic term should be half the total energy. Thus,
\[
 \frac{1}{2}\frac{p_{\phi}(v)^2}{R(v)^3}\sim
\frac{1}{2}R(v)^3\rho_{\rm matter}(v)
\]
and
\[
 H_{\rm quantum}= \frac{C}{2}\sum_{v}\rho_{\rm matter}(v)\ell_{\rm
P}^2R(v)\,.
\]
Finally, the patch size is the total size of the universe divided by
the number ${\cal N}$ of patches, $R(v)^3=a^3/{\cal N}$, resulting in
\[
H_{\rm quantum}= \frac{C}{2}\sum_v\rho_{\rm matter}(v)\ell_{\rm
  P}^2a/{\cal N}^{1/3}\sim\frac{C}{2}{\cal N}^{2/3}\rho_{\rm
  matter}\ell_{\rm P}^2a
\]
for a nearly homogeneous matter distribution among the ${\cal N}$
patches.

This is the contribution obtained by adding up quantum corrections for
all ${\cal N}$ patches. Since ${\cal N}$ is increasing with $a$,
$H_{\rm quantum}$ indeed contributes negative pressure resulting from
an energy contribution increasing with volume. More precisely, for the
case\footnote{We normalize the scale factor such that ${\cal N}={\cal
    N}_0$ for $a=\ell_{\rm P}$ is the number of vertices of a Planck
  size universe, which one typically expects to be of order one. This
  implies that the current scale factor in this normalization is
  huge.} ${\cal N}(a)\sim {\cal N}_0(a/\ell_{\rm P})^3$ with some
constant ${\cal N}_0$ of order one, we obtain
\[
 H_{\rm quantum}\sim \Lambda a^3
\]
with a (cosmological) constant $\Lambda$. This implies an equation of
state where pressure is the negative energy density, the currently
preferred value by observations. Moreover, the size of
$\Lambda=\frac{1}{2}C{\cal N}_0^{2/3}\rho_{\rm matter}$ is the same as
that of matter up to constant factors of order one. This case
corresponds to $\alpha=1$ in the parameterization ${\cal N}(V)\propto
V^{\alpha}$ as a function of volume motivated in Sec.~\ref{s:Ham}. As
explained there, the balance between changing edge spins and the
creation of new vertices provides this value as a limiting case of a
general range $0<\alpha<1$. For $\alpha\not=1$, we have ${\cal
N}(a)\sim {\cal N}_0(a/\ell_{\rm P})^{3\alpha}$, resulting in
\begin{equation}
  H_{\rm quantum}\sim \frac{C}{2}{\cal N}^{2/3}\rho_{\rm matter}\ell_{\rm P}^2a
  =\lambda a^{1+2\alpha}
\end{equation}
with a constant $\lambda=\frac{1}{2}C{\cal N}_0^{2/3}\rho_{\rm matter}
\ell_{\rm P}^{2(1-\alpha)}$.  In this case, the energy density of the
quantum correction ``fluid'' is
\[
 \rho_{\rm quantum}= a^{-3}H_{\rm quantum}\sim \lambda a^{2(\alpha-1)}
\]
and its pressure (defined as the negative derivative of energy by volume)
\[
 P_{\rm quantum} = -\frac{\md H_{\rm quantum}}{\md (a^3)}=
 -\frac{1+2\alpha}{3} \lambda a^{2(\alpha-1)}= -\frac{1+2\alpha}{3}
 \rho_{\rm quantum}\,.
\]
The equation of state parameter is thus
\begin{equation} \label{w}
 w=-\frac{1+2\alpha}{3}
\end{equation}
which is between $-1/3$ and $-1$ for the range $0<\alpha<1$.

Qualitatively, we rather naturally obtain all the observational
properties without additional input. Although the precise form of
lattice refinements or a precise estimate for $\alpha$ is not yet
available, falsifiable predictions are possible. In particular, one
can relate the equation of state parameter to the size of the current
contribution by dark energy. If $\alpha=1$, as we have seen, we obtain
$w=-1$ and a value of the cosmological constant of the same order as
the matter energy density. The Planck length dropped out of the final
expression, and thus there is no remaining factor of $a/\ell_{\rm P}$
in relating $\rho_{\rm quantum}$ to $\rho_{\rm matter}$, which could
give large or small factors compared to the matter density. If such
factors were present, they would result in a cosmological constant off
by orders of magnitude unless they appear in a form taken only to
small powers. If $\alpha\not=1$, the Planck length does not cancel out
completely and one has to worry about large factors in the final value
for $\rho_{\rm quantum}$ not of the same order as the matter density.
Ratios of $a/\ell_{\rm P}$ appear in 
\begin{equation} \label{l}
  \frac{\rho_{\rm quantum}}{\rho_{\rm matter}} = 
\frac{1}{2}C{\cal N}_0^{2/3} 
\left(\frac{\ell_{\rm P}}{a}\right)^{2(1-\alpha)}
\end{equation}
with small powers only if $\alpha$ is near one. For $\alpha\approx1$
one still obtains a dark energy contribution of the order of the
matter density, giving a phenomenological constraint on the value for
$\alpha$. Even if $\alpha$ cannot be computed easily from the
fundamental gravitational Hamiltonian to confirm that a value near
$\alpha=1$ is indeed realized, there is an interesting
phenomenological cross-check: As we have seen, the value of $\alpha$
also influences the equation of state of dark energy. Near $\alpha=1$
the equation of state is close to that of a cosmological constant,
$w=-1$. For smaller values of $\alpha$ one approaches the equation of
state parameter $w=-1/3$. Thus, the scenario described here, giving a
relation between the equation of state parameter and the order of
magnitude of dark energy density in terms of a single parameter
$\alpha$, is testable by precise observations of the dark energy
parameters, including its equation of state.

\section{Conclusions}

Loop quantum gravity becomes better and better understood and its
crucial properties are studied in more and more realistic
models. Since recently, inhomogeneous situations are being looked at
in detail and the basic mathematical framework has been derived. This
has led to several possible ingredients to solving the dark energy
problem, although no complete proposal for a solution exists so far.
Well-defined matter Hamiltonians make it possible at least in
principle to determine what the size of vacuum energy in a
semiclassical state is, and how it couples to gravity to play the role
of dark energy. While explicit calculations at a fundamental level are
still out of reach, a framework for effective equations applicable to
this situation exists. In this way, several technical and conceptual
difficulties can be circumvented. For instance, one would not need
completely specified semiclassical states but can systematically
determine their properties order by order in a semiclassical
expansion. Also this is still being worked out, and so no clear result
concerning dark energy exists. What does exist are phenomenological
models as described here which allow one to test if typical effects of
quantum gravity can have implications even in a large, classical
universe and contribute to the understanding of dark energy.

The same basic ingredients which in isotropic models lead to
inflationary behavior can be seen to imply properties of dark energy
even in a large universe. Although these applications come from the
same basic expression, different corrections are being probed. In
isotropic models of the early universe main effects arise from
non-perturbative corrections, while the effects used here for dark
energy rely on perturbative ones. Moreover, dark energy effects are
much more sensitive to properties of the full quantum dynamics,
especially the occurrence of lattice refinements which does not play a
large role in the early universe. Some of the effects present in
isotropic models have been shown to be dual to versions of string or
brane-world cosmology \citep{JimDual,BraneDual,RSLoopDual}. An
extension to inhomogeneous situations might provide interesting
comparisons between the different frameworks.

A successful implementation of a scenario describing dark energy
depends on several different aspects which allow qualitative
evaluations of the viability of effects: First and foremost, the main
correction arises in the kinetic term.  Matter components in this term
always appear in positive combinations such as $p_{\phi}^2$ for a
scalar, and quantum corrections generally contribute a positive
quantity since $C>0$ in (\ref{da}). This implies that small
corrections in patches can indeed add up coherently in such a way that
they produce positive dark energy. Dark energy could thus provide an
example for magnification effects of tiny quantum corrections by
adding them up over many contributions all over space. A similar
effect, where corrections add up during long evolution times, may
leave observable imprints in the CMB \citep{InhomEvolve}.  Other
crucial ingredients are the kinds of basic variables used, which
affects the $a$-dependence of quantum corrections, and the discrete
structure. All this results in corrections which are not yet uniquely
determined from the theory but whose qualitative properties can be
estimated up to constants of order one.

Qualitative explanations which are realized non-trivially are: (i) the
subdivision into patches as it follows from technical expectations
agrees with the observed range for the equation of state of dark
energy; (ii) the size of $\Lambda$ is automatically related to the
energy density of matter because it results from quantum corrections
to its kinetic term; (iii) there is a tight relation between the size
and equation of state of dark energy, Eqs.~(\ref{l}) and (\ref{w}).
These observations are not independent but closely related to each
other: if functions ${\cal N}(a)$ not close to cubic behavior had
occurred, powers of $a/\ell_{\rm P}$ had appeared in $\rho_{\rm
quantum}/\rho_{\rm matter}$ which are far from being of order
one. While still related to $\rho$, $\Lambda$ would then have been too
big ($\alpha>1$) or too small ($\alpha<1$) by many orders of
magnitude. That the technical expectation agrees with the observation
in this way is a non-trivial feature.

The scenario is thus constrainable and testable by more detailed
observations of cosmic acceleration, as well as structure formation
which probe different parts of the theoretical curve $d(a)$. This has
to be combined with detailed calculations and a dedicated effort in
numerical quantum gravity to simulate the subdivision better and to
obtain precise predictions. While some numerical aspects have occurred
in loop quantum gravity, this was so far only in homogeneous models
\citep{Time,APS}. Inhomogeneous ones as needed for dark energy require
much more refined methods. Interestingly, though, current
investigations of the semiclassical behavior of homogeneous models in
loop quantum cosmology indicate that a value near $\alpha=1$ is
preferred, while values near $\alpha=0$ would not provide the correct
semiclassical limit \citep{APSII,SchwarzN}. This can be seen as
circumstantial evidence for the dark energy scenario proposed here,
which is to be corroborated by dedicated studies of inhomogeneous
models.  It may be worth the effort because a detailed knowledge of
${\cal N}(a)$ would allow one to predict the far future of the
universe from such models: If an analysis confirms that ${\cal
  N}(a)\sim a^3$, the patch size remains nearly constant during
expansion and there will be no dramatic changes in the future
evolution. But if ${\cal N}$ increases by a different power, the
equation of state will change. Moreover, if the power becomes smaller
than three, the patch size increases with expansion and dark energy
will die off. If the power exceeds three, the patch size decreases and
slowly approaches the strongly modified non-perturbative part of
$d(a)$. At this point, the behavior of the universe would change
dramatically through quantum effects on large scales. This case would
require the edge spins to decrease systematically while the total
volume grows which is, for better or worse, not supported by current
constructions of loop quantum gravity.

\section*{Acknowledgements}

This work was supported in part by NSF grant PHY 05-54771.


\end{document}